\documentclass[conference]{IEEEtran}
\IEEEoverridecommandlockouts
\usepackage{cite}
\usepackage{amsmath,amssymb,amsfonts}
\usepackage{algorithm2e}
\usepackage{graphicx}
\usepackage{textcomp}
\def\BibTeX{{\rm B\kern-.05em{\sc i\kern-.025em b}\kern-.08em
    T\kern-.1667em\lower.7ex\hbox{E}\kern-.125emX}}
    
\begin{document}

\title{Inferring users' preferences through leveraging their social relationships\\
\thanks{This work is supported by the National Natural Science Foundation of China (Grant Nos. 11622538, 61673150, 71662014, 61403114 and 71661015). L.L and Chunxiao Jia acknowledge the Zhejiang Provincial Natural Science Foundation of China (Grant Nos. LR16A050001 and LQ14F030009). X.F.D acknowledges the Jiangxi Provincial Science and Technology Program(Grant No. GJJ160292), Jiangxi Provincial Social Science Planning Project (Grant Nos. 16GL08, 16GL09) and the Jiangxi Provincial Humanities and Social Sciences Research Project (Grant No. JC1543). X.L.R acknowledges the support from China Scholarship Council (CSC). The preliminary results have been presented in Journal of Shandong University (Natural Science) in Chinese. Correspondence should be addressed to L.L (linyuan.lv@gmail.com).}
\author{\IEEEauthorblockN{
Xiaofang Deng\IEEEauthorrefmark{2},\IEEEauthorrefmark{1}
Leilei Wu\IEEEauthorrefmark{1},
Xiaolong Ren\IEEEauthorrefmark{3},
Chunxiao Jia\IEEEauthorrefmark{4},
Yuansheng Zhong\IEEEauthorrefmark{5},
and
Linyuan L\"{u}\IEEEauthorrefmark{1},\IEEEauthorrefmark{4}}
\IEEEauthorblockA{\IEEEauthorrefmark{1}Institute of Fundamental and Frontier Sciences\\
University of Electronic Science and Technology of China, Chengdu 610054, China \\
Email: wuleilei868@gmail.com; linyuan.lv@gmail.com}
\IEEEauthorblockA{\IEEEauthorrefmark{2}School of Software, Jiangxi Normal University, Nanchang 330022, China\\ 
Email: dxf@jxnu.edu.cn}
\IEEEauthorblockA{\IEEEauthorrefmark{3}Computational Social Science, ETH Zurich, Zurich CH-8092, Switzerland\\
Email: xren@ethz.ch}
\IEEEauthorblockA{\IEEEauthorrefmark{4}Alibaba Research Center for Complexity Sciences, Hangzhou Normal University, Hangzhou 311121, China\\
Email: chunxiaojia@163.com}
\IEEEauthorblockA{\IEEEauthorrefmark{5}College of Information Technology, Jiangxi University of Finance and Economics, Nanchang 330013, China\\
Email: zhong.ys@163.com}}}
\maketitle
\begin{abstract}
Recommender systems, inferring users' preferences from their historical activities and personal profiles, have been an enormous success in the last several years. Most of the existing works are based on the similarities of users, objects or both that derived from their purchases records in the online shopping platforms. Such approaches, however, are facing bottlenecks when the known information is limited. The extreme case is how to recommend products to new users, namely the so-called cold-start problem. The rise of the online social networks gives us a chance to break the glass ceiling. Birds of a feather flock together. Close friends may have similar hidden pattern of selecting products and the advices from friends are more trustworthy. 

In this paper, we integrate the individual's social relationships into recommender systems and propose a new method, called Social Mass Diffusion (SMD), based on a mass diffusion process in the combined network of users' social network and user-item bipartite network. %The results show that the new algorithm can not only improve the prediction accuracy for inactive users, but also increase the recommended diversity for active users.??
The results show that the SMD algorithm can achieve higher recommendation accuracy than the Mass Diffusion (MD) purely on the bipartite network. Especially, the improvement is striking for small degree users. Moreover, SMD provides a good solution to the cold-start problem. The recommendation accuracy for new users significantly higher than that of the conventional popularity-based algorithm. These results may shed some light on the new designs of better personalized recommender systems and information services.
\end{abstract}

\begin{IEEEkeywords}
Recommendation; Social networks; Mass diffusion algorithm; Cold-start problem
\end{IEEEkeywords}

\section{Introduction}
In the big data era, people are often faced with the information overload problem. How to efficiently find the useful information is a challenging problem. Recommender systems \cite{LY'12} are a subclass of an information filtering system that helps users to find what they really want in the larger-scale online shop and online retailers. Many diverse recommendation techniques have been developed, including content-based analysis \cite{Ansari'00, Pazzani'07, Adomavicius'05}, collaborative filtering \cite{GNOT'92,HKTR'04,LL'10}, spectral analysis \cite{Maslov'01}, and the network-based recommender systems \cite{Zhang'07,Zhou'07,ZKLMW'10}. Despite all of these efforts, recommender systems still face many challenges. There are demands for further improvements on the prediction accuracy and the recommended diversity. Meanwhile, there is also the cold-start problem \cite{SPU'02}, that is how to accurately recommend for new users or users with few records. 

Due to the fact that people are more likely to be friends if they have more common interests, in recent years, many applications on recommendation well utilized users' social relationships to improve recommendation accuracy, such as social friends recommender systems (Facebook), social restaurant recommender systems (Yelp and Dianping), social movie recommender systems (Flixster), and social book recommender systems (Shelfari and Douban). The success of these applications demonstrates that taking into consideration social connections can improve the performance of recommender systems. Although some social-based recommendation algorithms have been proposed \cite{Li'15,Nie'14,Xu'15,Jian'10,Ralf'08,Ma'08,Shepitsen'08}, how to devise an algorithm of both effectiveness and efficiency is still challenging. 

Diffusion-based algorithms on user-item bipartite networks are well-known for the simplicity, such as heat conduction methods \cite{Zhang'07,ZhouYB'13}, mass diffusion methods \cite{Zhou'07,LL'11} and the hybrid of the two \cite{ZKLMW'10}. In this paper, we integrate the users' social network and the user-item bipartite network, and consider the mass diffusion process on the combined network to develop a new recommendation algorithm called Social Mass Diffusion (SMD). Two datasets, namely \textit{FriendFeed} \cite{CDM'10} and \textit{Epinions} \cite{MA'06} are used to test the algorithm's performance. The results show that social information can help to improve the recommendation accuracy especially for inactive users. The algorithm also serves as a good candidate to replace the conventional popularity-based algorithms to provide more personalized recommendations in the cold-start period.

\section{The model}\label{Model}
\begin{figure}[htbp]
\centering
\begin{tabular}{ccc}
\includegraphics[scale = 0.16]{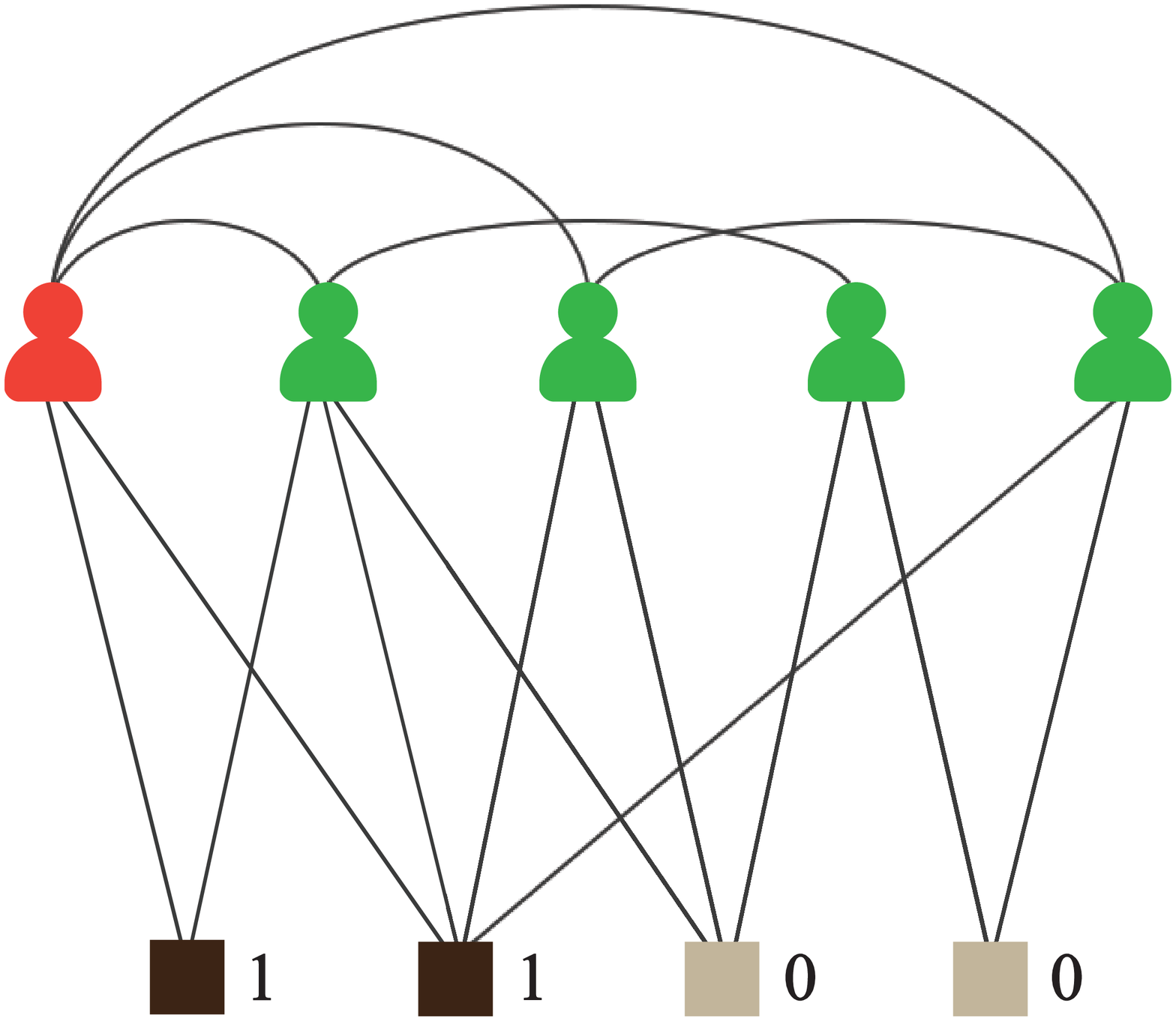} &
\includegraphics[scale = 0.16]{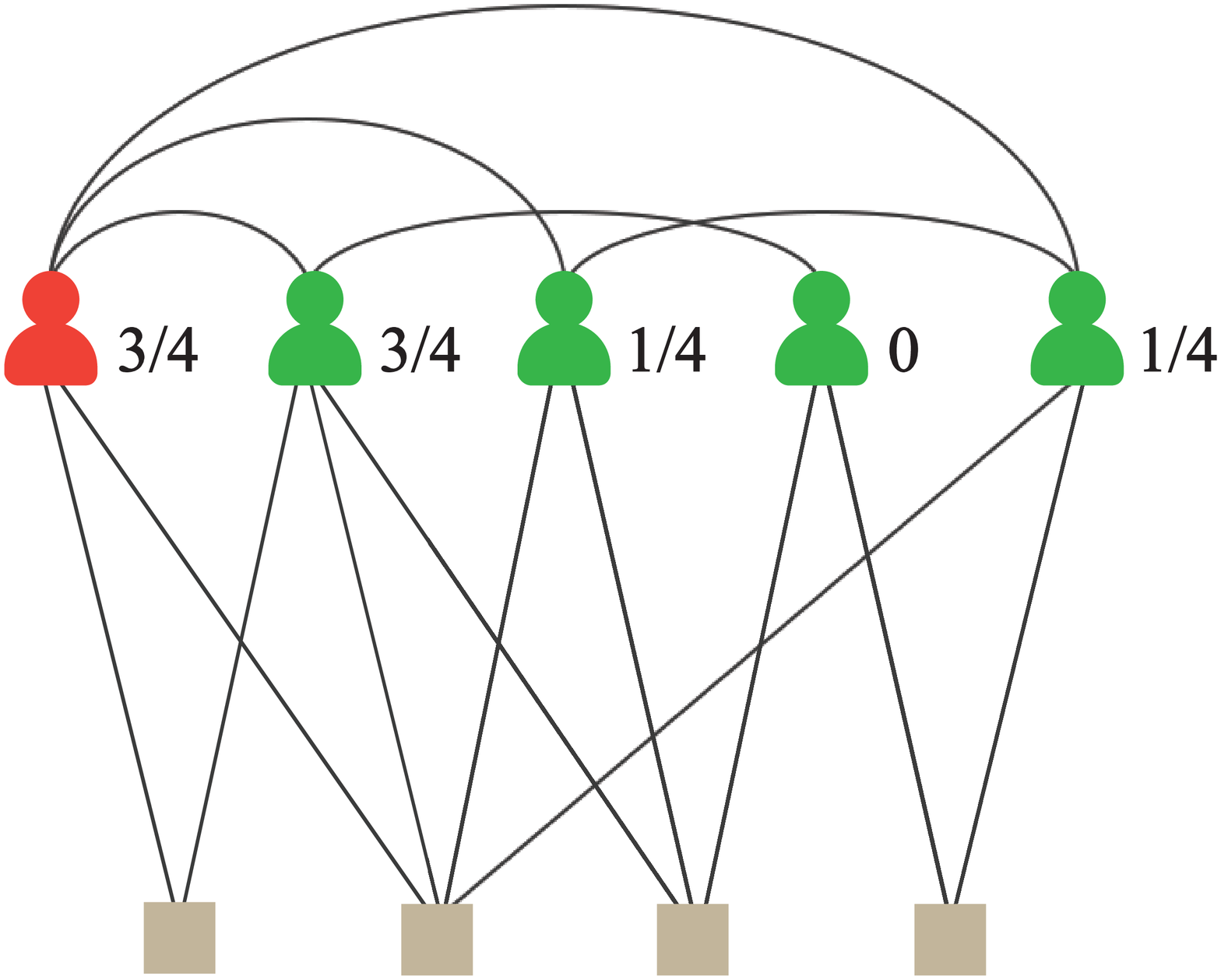} &  \\
(a) Initialization & (b) First-step diffusion&  \\
\includegraphics[scale = 0.16]{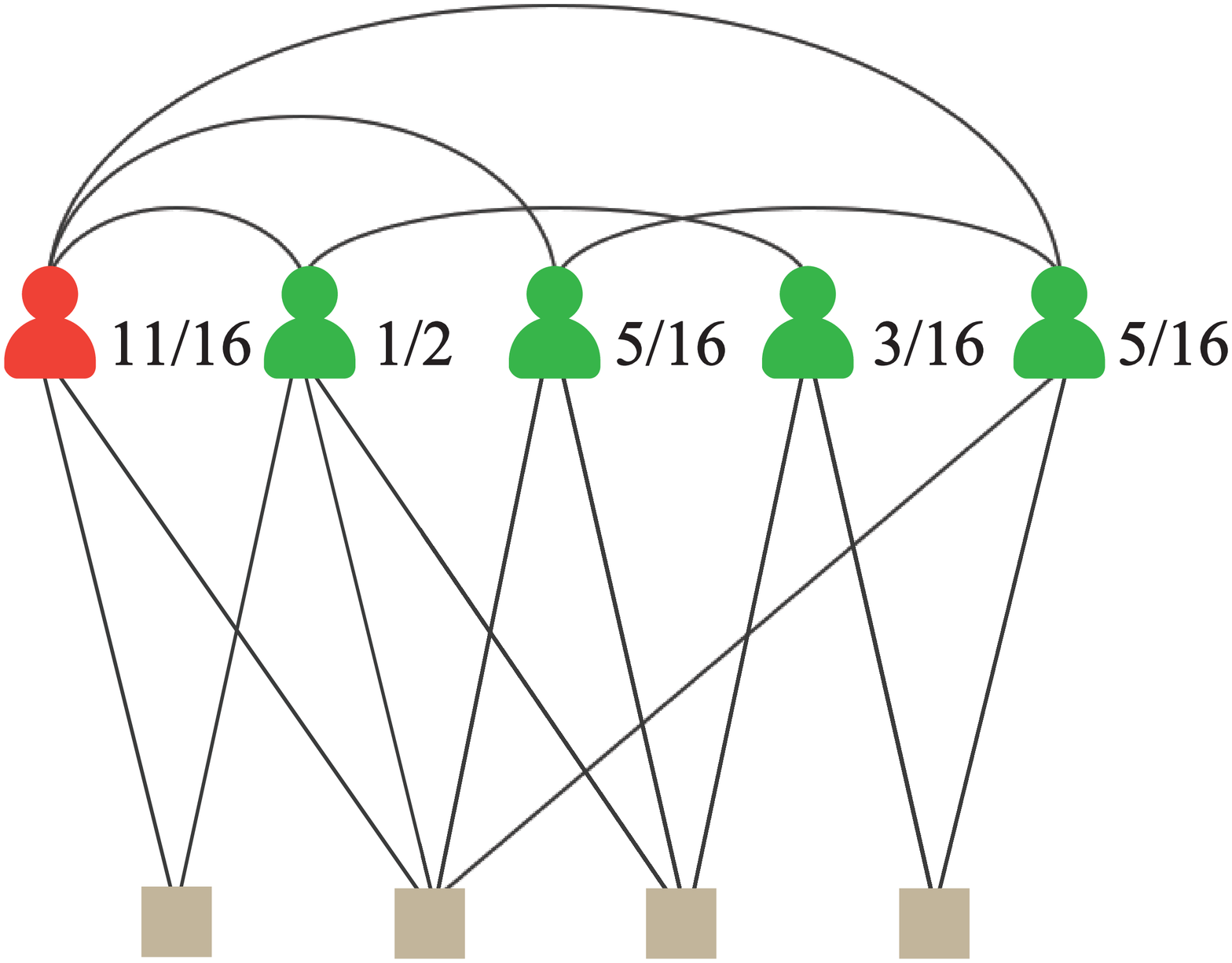} &
\includegraphics[scale = 0.16]{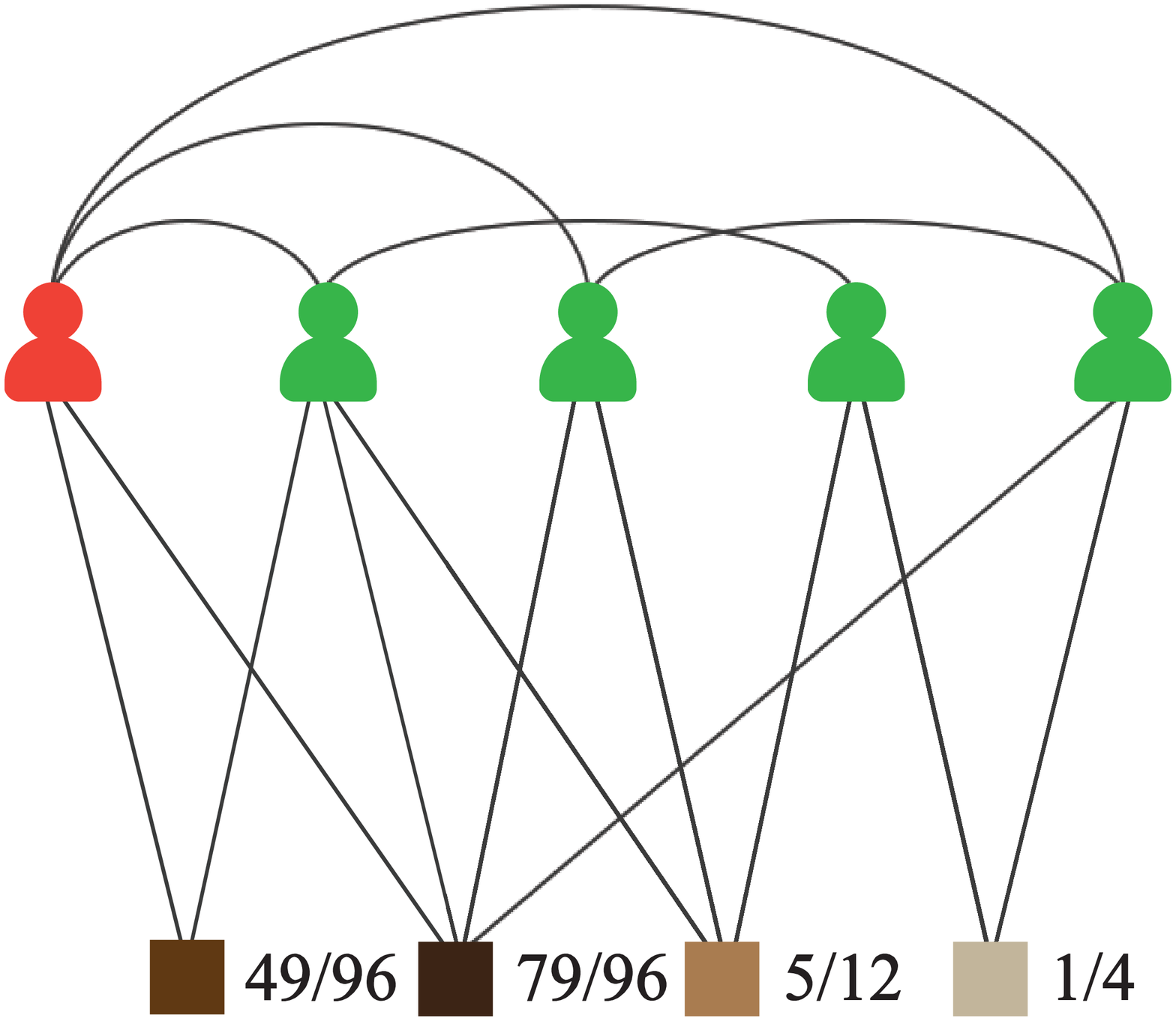} &  \\
(c) Social diffusion & (d) Final diffusion&  \\
&  &
\end{tabular}%
\caption{An illustration of the SMD method with $p=0.5$. The user in red is the target user, and squares denote items. The darker the squares, the greater the amount of resources they own. The numbers indicate the amount of resources the users or the items have.}
\label{Sketchmap}
\end{figure}

We consider a user-user-item network denoted by $G(U,O,E_{UO},E_{UU})$, where $U$ denotes a group of $n$ users, $O$ denotes a group of $m$ items, $E_{UO}$ denotes the set of edges between users and items, and $E_{UU}$ denotes the set of edges between users. Actually, the network is a combination of two sub-networks, namely a user-item bipartite network $G_{UO}(U,O,E_{UO})$ and a user-user social network $G_{UU}(U,E_{UU})$ (as shown in figure~\ref{Sketchmap}(a)). The user-item bipartite network $G_{UO}$ can be characterized by an $n\times m$ adjacency matrix ${\cal A}_{UO}$, where the matrix element $a_{i\alpha }=1$ if user $i$ has collected item $\alpha$, and otherwise $a_{i\alpha }=0$. Similarly, the social network $G_{UU}$ can be characterized by an $n\times n$ adjacency matrix ${\cal S}_{UU}$, where the matrix element $s_{ij}=1$ if user $i$ and $j$ are friends, and otherwise $s_{ij}=0$. For the user-item bipartite network, we denote the number of users who have collected item $\alpha $ to be $k_\alpha$, and number of items that have been collected by user $i$ to be $k_i$. Hence, the $k_\alpha$ and $k_i$ are the degree of item $\alpha$ and user $i$ in the user-item bipartite network, respectively. On the other hand, we denote the number of friends of user $i$ in the social network to be $K_i$, which we call the social degree of user $i$. 

\subsection{Classic diffusion-based recommendation methods}
In this section, we introduce three classic diffusion-based recommendation methods, namely Mass Diffusion (MD) method \cite{Zhou'07}, Heat Conduction method (HC) \cite{Zhang'07} and the Hybrid method \cite{ZKLMW'10}. Many variants and improvements of the originally algorithms have also been proposed recently ~\cite{LY'12,ZhouYB'13,RLL'14,YZGM'16}. 

\textbf{Mass diffusion method (MD).} This diffusion-based method was originally proposed by considering the resource-allocation process on user-item bipartite network \cite{Zhou'07}. It works by assigning one unit of resource on each of the items collected by the target user, and subsequently spreading the resources on the user-item bipartite network to the items potentially favored by the target user. Mathematically, the initial resources on all the items can be denoted by a vector $\vec{f}$ where $f_\alpha=1$ if item $\alpha$ is collected by the target user, otherwise $f_\alpha=0$. Then the diffusion process is described by the equation $\vec{f^{^{\prime }}}=W_{\rm MD}\vec{f}$, where $\vec{f^{^{\prime }}}$ is the final resources allocated to the items after two-steps mass diffusion process. If the resources are equally distributed in every steps, the transfer matrix $W_{\rm MD}$ is an $m\times m$ matrix with the element $w_{\alpha \leftarrow \beta}^{\rm MD}$ given by
\begin{equation}
 w_{\alpha \leftarrow \beta}^{\rm MD}=\dfrac{1}{k_{\beta }}\sum_{i=1}^{n}\dfrac{
a_{i\alpha }a_{i\beta }}{k_{i}}.
\end{equation}
After the two-steps mass diffusion process, items that not yet collected by the target user are ranked according to their resources in $\vec{f^{^{\prime }}}$ in a descending order. The top-ranked items are assumed have higher probabilities to be preferred by the target user, and will be recommended accordingly.

\textbf{Heat conduction method (HC).} Heat conduction method~\cite{ZKLMW'10} is inspired by the heat diffusion process on an user-item bipartite network, in which all the items selected by the target user are considered as a stationary heat source while other items are cold points. After a long time, the cold points will obtain a certain amount of heat. Comparing to equally distributing the resource to the nearest neighbors in MD method, the amount of heat possessed by every uncollected item equals to the average temperatures carried by its neighbors in HC method. Mathematically, the transfer matrix $W_{HC}$ is an $m\times m$ matrix with the element $w_{\alpha \leftarrow \beta}^{\rm HC}$ given by
\begin{equation}
 w_{\alpha \leftarrow \beta}^{\rm HC}=\dfrac{1}{k_{\alpha }}\sum_{i=1}^{n}\dfrac{
a_{i\alpha }a_{i\beta }}{k_{i}}.
\end{equation}

\textbf{MD-HC hybrid method (Hybrid).} In general, MD method is good at obtaining accurate recommendation lists, while HC method is powerful to gain more diverse recommendation lists. In order to solve the diversity-accuracy dilemma, an elegant hybrid solution is proposed by introducing a tunable parameter into the transition matrix normalization~\cite{ZKLMW'10}:
\begin{equation}
w_{\alpha \leftarrow \beta}^{\rm Hybrid}= \frac{1}{k^{1-\lambda}_{\alpha}k^{\lambda}_{\beta}}\sum_{i=1}^{n}\frac{
a_{i\alpha }a_{i\beta }}{k_{i}}.
\end{equation}

\subsection{Social mass diffusion model (SMD)}
Recommendations derived from the above diffusion process are merely based on the users' historical collecting records without considering their social relationships. However, in real life, when we want to buy a product that is not familiar, we often consult with our friends who have already had experiences with the product. We are always likely to adopt the recommendations from our friends rather than the salesman. By considering the effects of social relationships on users' shopping behaviors, we propose a social mass diffusion method. The resources are distributed not only on the user-item network but also the user-user friendship network. There are four steps in the SMD method: (i) Initialization: Every item collected by the target user is assigned one unit of resources. (ii) First-step diffusion: The resources are equally distributed to users who have collected them. (3) Social diffusion: Each user keep a ratio $p$ of his/her obtained resources, and the rest $1-p$ resources will be evenly distributed to his/her friends (i.e., one-step diffusion on user-user social network). Note that, we can also consider a multi-steps diffusion. (iii) Final diffusion: A user's total resources, including the reserved $p$ part and those come from his/her friends, are also averagely distributed to his/her purchased items. Then items not yet collected by the target user are ranked according to their resources in a descending order, and the top-ranked ones will be recommended. Mathematically, the transition matrix $W_{SMD}$ reads
\begin{equation}
w^{SMD}_{\alpha \leftarrow \beta }=p\dfrac{1}{k_{\beta }}\sum_{i=1}^{n}\dfrac{
a_{i\alpha }a_{i\beta }}{k_{i}}+(1-p)\dfrac{1}{k_{\beta }}
\sum_{i=1}^{n}\sum_{j=1}^{n}\dfrac{a_{i\alpha
}A_{ij}a_{j\beta }}{k_{i}K_{j}},
\end{equation}
where the first term on the right side describes the diffusion process merely on the user-item bipartite network, while the second term describes the diffusion process goes through the social network. If $p=1$, SMD method degenerates to the original MD method. If $p=0$, all resources will go through the social networks. Note that, SMD method also works for new users where MD fails, and thus solves the cold-start problem. If the target user has not purchased any item, SMD starts by assigning one unit of resource to the target user, and the resource goes through the social network and finally diffuses to the item side. Figure ~\ref{Sketchmap} gives an illustration of the SMD method with $p=0.5$.

\section{Evaluation metrics} \label{DataMetrics}
To evaluate the performance of the algorithm, we divide all links in the user-item bipartite network into two sets randomly: the training set $E^{T}$\ which contains $90\%$ of all links, and the remaining $10\%$ are considered as the probe set $E^{P}$. We employ seven metrics to fully characterize the algorithm's performance:

(i) \textbf{Ranking Score(\textit{RS})\cite{Zhou'07}: }Ranking Score measures the ability of a recommendation algorithm to produce a good ordering of items that matches the user's preference. For a target user, the recommender system returns a list of all his uncollected items ranked according to his preference. For each user-item link in the probe set, we measure the rank of the items in the recommendation list of the users. A good algorithm is expected to give those items a higher rank, and thus leads to a small \textit{RS}. Averaging over all the users' ranking scores in the probe set, we obtain an average ranking score \textit{RS} that quantify recommendation accuracy of the algorithm. Alternatively, \textit{RS} is given by
\begin{center}
$ \textit{RS} = \dfrac{1}{n}\underset{i}{\sum}\textit{RS}_{i}$,
\end{center}
where $\textit{RS}_{i}$ is the average ranking score of user $i$. Clearly, the smaller the ranking score, the higher the accuracy of the algorithm, and vice versa.

(ii) \textbf{Precision(\textit{P})\cite{HKTR'04}:} Precision is defined as the
ratio of relevant items selected by a target user to the number of items found in the top-$L$ positions of the recommendation list. For a target user $i$, the precision $P_{i}(L)$ is defined as
\begin{center}
$\qquad P_{i}(L)=\dfrac{d_{i}(L)}{L}$,
\end{center}
where $d_{i}(L)$ is the number of relevant items (i.e., the items
collected by $i$ in the probe set) in the top-$L$ positions of the recommendation list. By averaging the individual precisions over all users with
at least one link in the probe set, we obtain the mean precision $P(L)$ of the whole
system. 

(iii) \textbf{Inter-user diversity  of recommended products(\textit{H})\cite{ZJS'08}:} it measures the diversity of recommendation lists among different users.
Given two users $i$ and $j$, the difference of their lists can be
measured by the hamming distance:
\begin{center}
$H_{ij}(L)=1-\dfrac{C_{ij}(L)}{L}$,
\end{center}
where $C_{ij}(L)$ denotes the number of common items in the top-$L$ positions
of their lists. Obviously, $H_{ij}(L)=0$ if $i$ and $j$ obtain an identical recommendation list, while $H_{ij}(L)=1$ if their lists are completely
different. By averaging $H_{ij}(L)$ over all pairs of users, we obtain the
average inter-user recommendation diversity $H(L)$ for the system. The greater the value of $H(L)$, the more personalized the recommendation list to individual users.

(iv) \textbf{Intra-user diversity of recommended products(\textit{I})\cite{ZSL'09}:}  it quantifies the similarities among the items within the recommendation list of an individual user. For a target user $i$, the recommendation list can be denoted by $%
\{o_{1},o_{2},...,o_{L}\}$, and the similarity of products in his/her recommendation list can be defined as:
\begin{center}
$I_{i}(L)=\dfrac{1}{L(L-1)}\underset{\alpha ,\beta \in O_{i}(L)}{\underset{%
\alpha \neq \beta }{\sum }}S_{\alpha \beta }^{o}$,
\end{center}
where $S_{\alpha \beta }^{o}$ is the similarity between items $\alpha$ and $\beta$, and $O_i(L)$ is the item found in the top-$L$ positions of the recommendation list of user $i$. Here we employ the widely used cosine similarity index \cite{SM'83} to measure similarity between items. Given two items $\alpha$ and $\beta$, their similarity is defined as:
\begin{center}
$S_{\alpha \beta }^{o}=\dfrac{1}{\sqrt{k_{o_{\alpha }}k_{o_{\beta }}}}%
\underset{l=1}{\overset{n}{\sum }}a_{l\alpha }a_{l\beta }$,
\end{center}
By averaging $I_{i}(L)$ over all users, we obtain the average intra-user recommendation diversity $I(L)$ for the whole system. A good algorithm should be able to cover a sufficient area of interest for an individual user and obtains a low intra-user recommendation diversity.

(v) \textbf{Coverage(\textit{Cov})\cite{LL'11}:} Coverage is the ratio of the number of distinct items included in all user's recommendation lists to the total number of items in the system. It can be defined as:
\begin{center}
$Cov=\dfrac{1}{m}\overset{m}{\underset{\alpha =1}{\sum }}\delta _{\alpha }$,
\end{center}
where $m$ is the total number of items in the system and $\delta _{\alpha }=1$
only if item $\alpha$ appears in the recommendation list of at least one user, otherwise $\delta _{\alpha }=0$.

(vi) \textbf{Novelty(\textit{N})\cite{ZLZZ'10}:} A good recommendation algorithm
should be able to identify niche or unpopular items that users are less likely to find through other ways but match their preferences. The metric novelty quantifies the
capability of an algorithm to generate novel and unexpected results. The
simplest way to calculate the novelty is to average the recommended times of
items, as
\begin{center}
$N_{i}(L)=\dfrac{1}{L}\underset{\alpha \in O_i(L))}{\sum }k_{\alpha
}$,
\end{center}
By averaging $N_{i}(L)$ over all users, we obtain the average popularity for the system.

(vii) \textbf{Congestion(\textit{C})\cite{RLL'14}:} Congestion occurs while a few
different items are recommended to numerous users. It can be quantified by
the famous Gini coefficient which was used to measure the inequality of
individual wealth distribution in an economy. Firstly, we rank items in
an ascending order according to the number of occurrences in the recommended lists
of all users. Then the Lorenz curve, denoted by $R(x)$, is the normalized
cumulative times of recommendation, and $x\in \lbrack 0,1]$ indicate the
normalized rank. Finally, congestion is defined as:
\begin{center}
$C=1-2\int_{0}^{1}R(x)dx$
\end{center}
Obviously, $C=0$ indicates that all items are recommended to users with the
same probability.

\section{Data sets and results}
\subsection{Empirical analysis}

\begin{figure}[htb]
\centering
\begin{tabular}{ccc}
\includegraphics[width=0.225\textwidth]{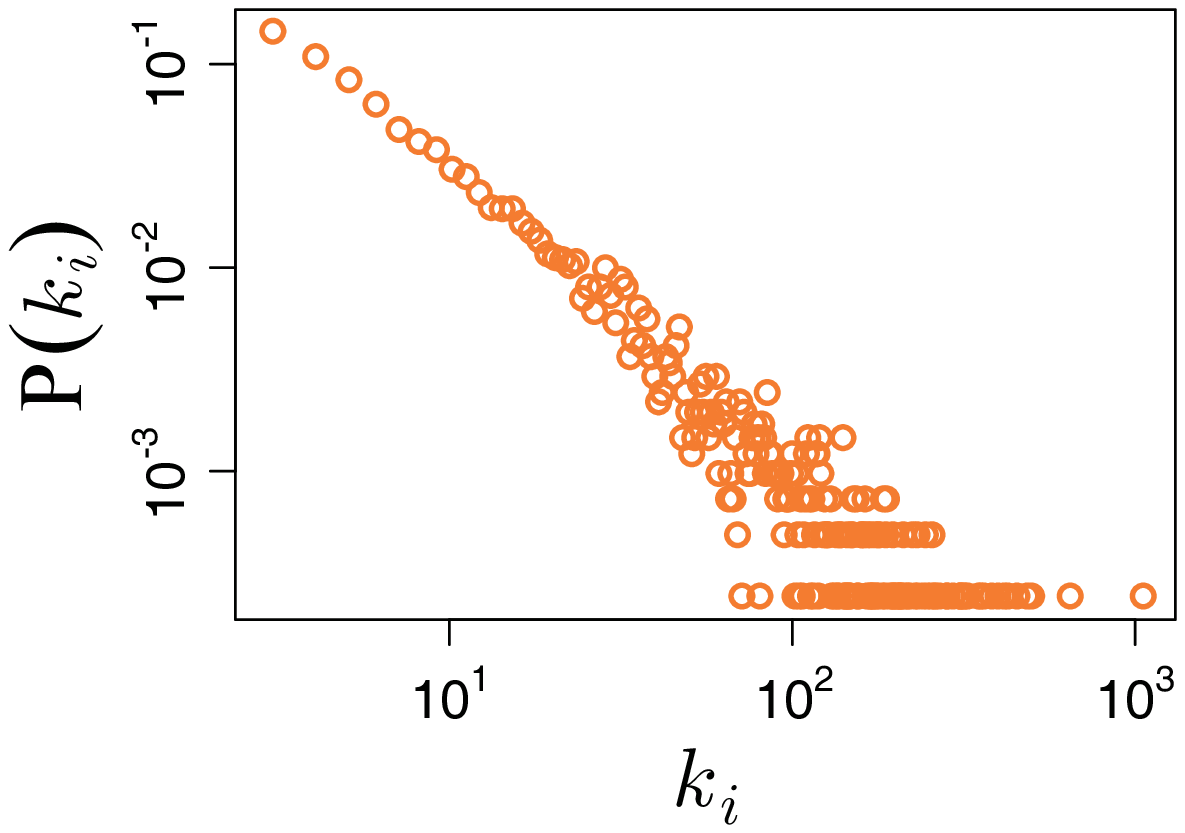} &
\includegraphics[width=0.225\textwidth]{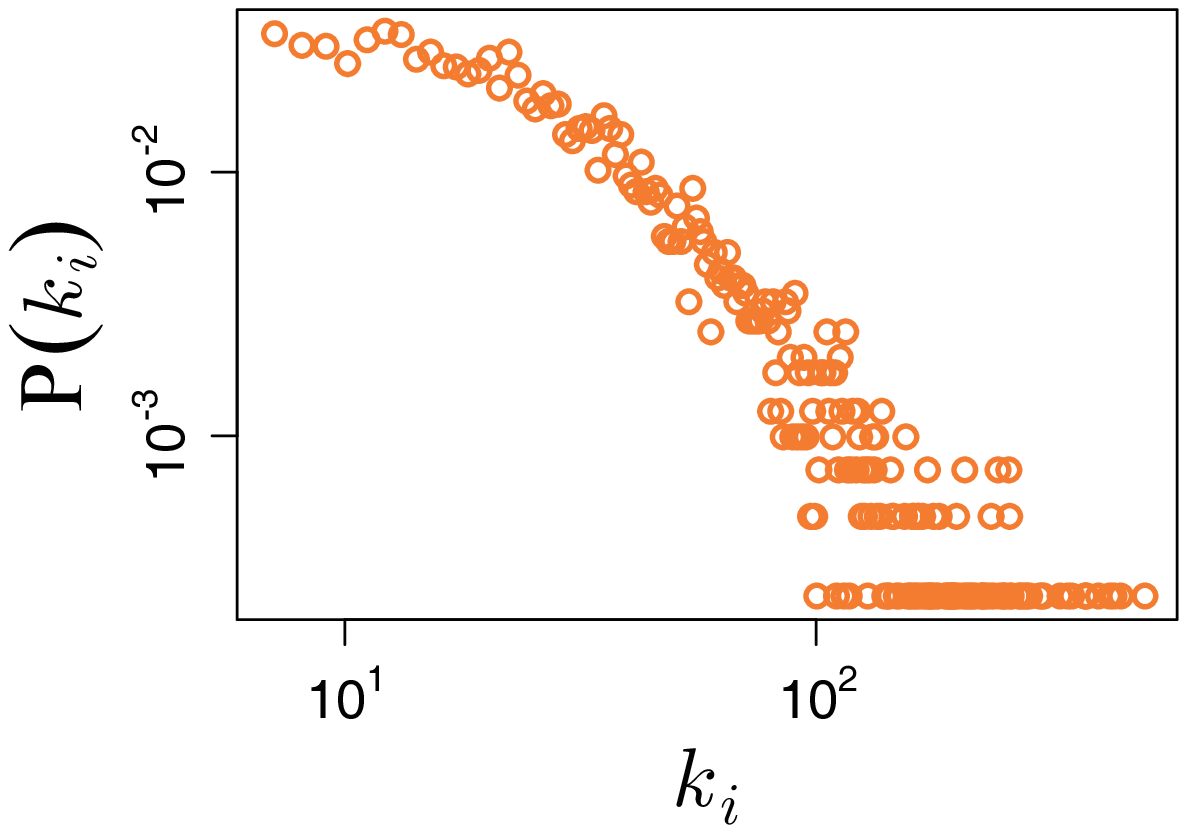} \\
(a) Friendfeed&(b) Epinions\\
\end{tabular}
\caption{The degree distribution of users in user-item bipartite network.}
\label{degreeDistribution}
\end{figure}

We consider two benchmark datasets, namely \textit{Friendfeed} \cite{CDM'10} and \textit{Epinions} \cite{MA'06}. \textit{Friendfeed} was a real-time feed aggregator that consolidates updates from social media and social networking websites. \textit{Epinions.com} was a general consumer review website established in \textit{1999}. Other than the conventional user-item relations, we can extract social relations among the users in these networks. Specifically, we consider user $i$ and user $j$ are friends whenever $i$ follows $j$ or $j$ follows $i$ in the original social network, resulting in an undirected social network. On the other hand, links in the user-item bipartite network are defined when a user collected an item. The statistical properties of the two datasets are given in table \ref{dataset_property}. From figure \ref{degreeDistribution}, we can see that the user degree $k_{i}$ in the user-item bipartite network follows a power-law-like distribution for both datasets. Comparing with \textit{Epinions}, there are more inactive users in \textit{Friendfeed}. The percentage of users with degree $k_{i}\leq 7$ are $44.46\%$ and $3.32\%$ in \textit{Friendfeed} and \textit{Epinions}, respectively. In addition, we find that if two users are friends, they have a higher probability to collect common items than strangers. There are $46.34\%$ ($54.5\%$) friend pairs in \textit{Friendfeed} (\textit{Epinions}) that co-collect at least one item. All these evidences support the fact that friends are more likely to collect items in common, and thus the social relationship information would be useful for inferring user preference in recommender systems.

\begin{table}[htbp]
\caption{The statistical properties of the two benchmark datasets.}
\label{dataset_property}\centering
\begin{tabular}{cccccccc}
\hline
Datasets & $n$ & $m$ & $|E_{UO}|$ & $|E_{UU}|$ & $\langle k_{i}\rangle$ & $\langle K_{i}\rangle$   \\
\hline
\textit{Friendfeed} & 4148 & 5700 & 96942 & 265497 & 23 & 128  \\
\textit{Epinions} & 4066 & 7649 & 154122 & 167717 & 37 & 82  \\ \hline
\end{tabular}%
\end{table}

\subsection{Performance of SMD method}

\begin{figure}[tbh]
\centering
\begin{tabular}{ccc}
\includegraphics[width=0.225\textwidth]{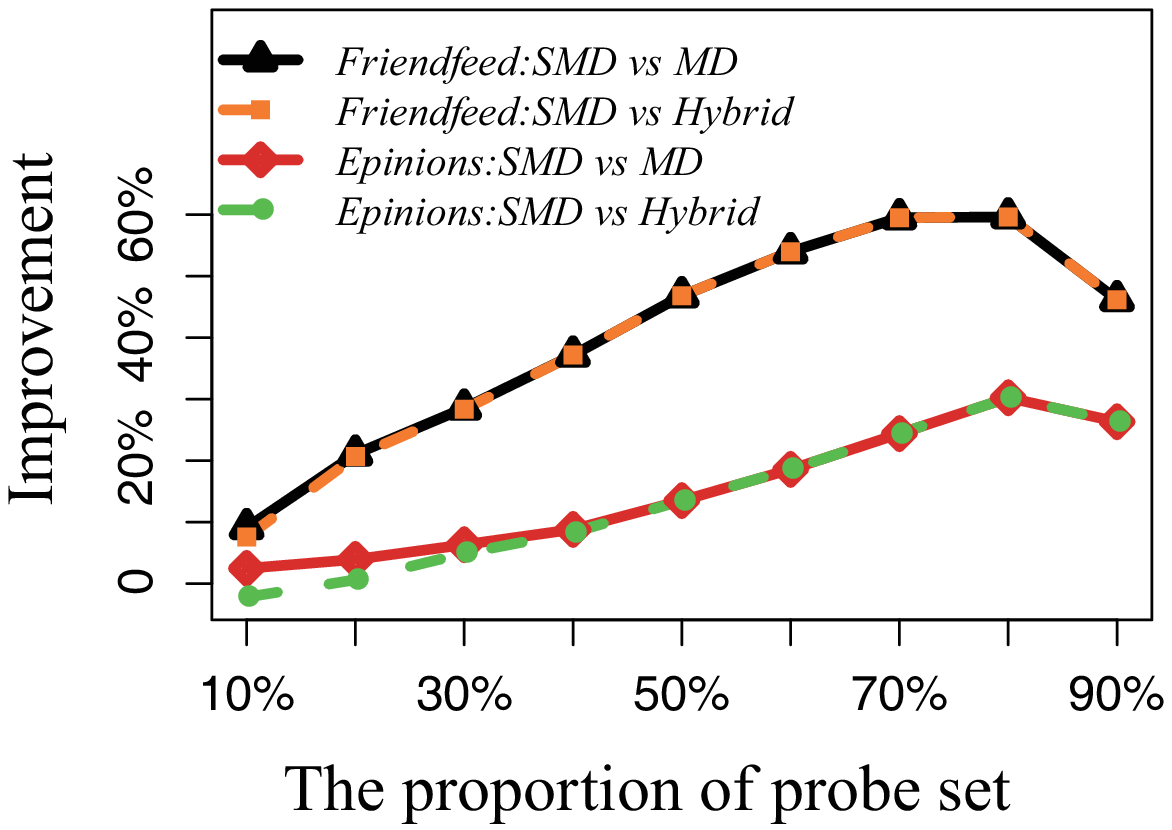} &
\includegraphics[width=0.225\textwidth]{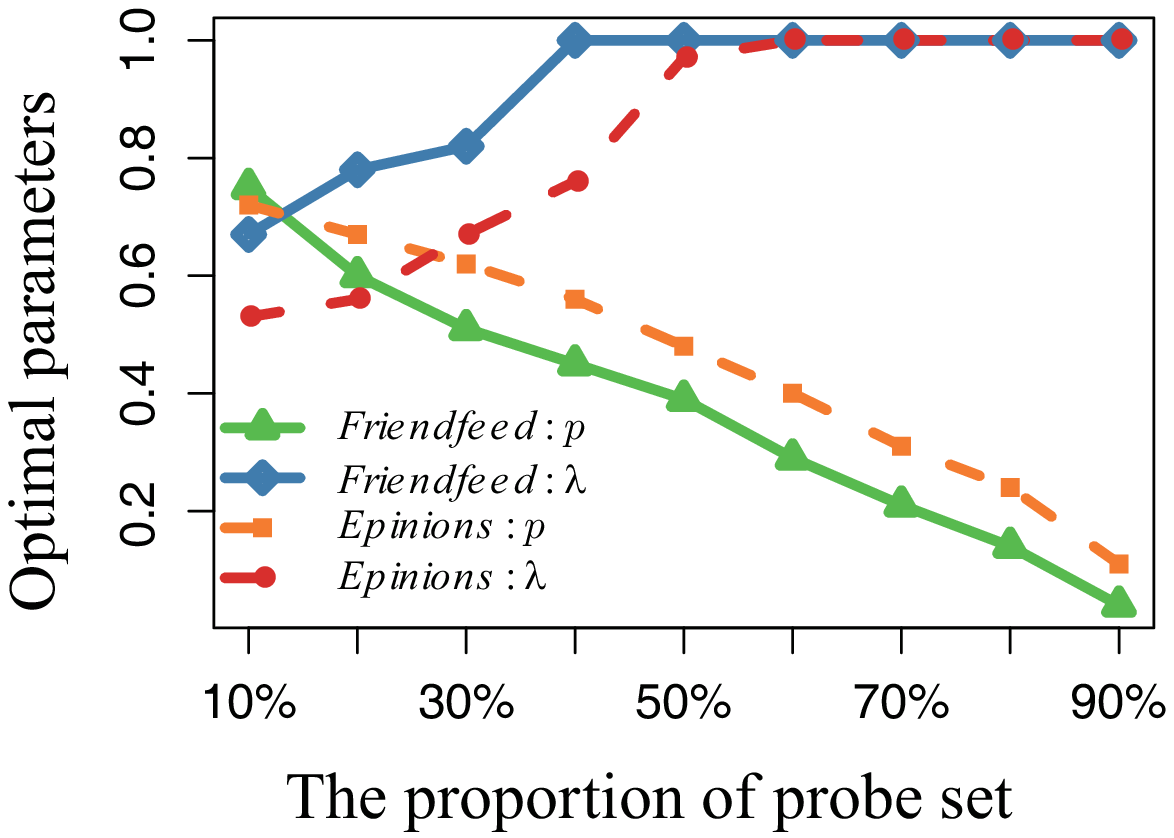} \\
(a) Improvement of SMD&(b) Optimal parameters\\
\end{tabular}
\caption{The accuracy improvements of SMD over MD and Hybrid methods. Each value is obtained by averaging over ten independent runs.}
\label{Proportion_ProbeSet}
\end{figure}

The performances of four methods on two datasets are shown in table \ref{compareWithOtherAlgorithms}. SMD outperforms all the previous diffusion methods on the \textit{Friendfeed} dataset in terms of ranking score, but performs less favorably compared with the hybrid method on \textit{Epinions}. This may be explained by the difference on the sparsity of the two datasets. \textit{Friendfeed} is denser than \textit{Epinions}, which implies that there is more social information for SMD to exploit in \textit{Friendfeed}. Among all metrics, the accuracy metric is of the most importance, since it directly shows the algorithms' capability to create commercial profits. Although the SMD is less effective to generate diverse and novel recommendations, we can also improve the performances on these aspects by turning the parameter $p$. For example, when $p=0$, that is, all the resources are spreading along the social relationships, in the first-step diffusion of SMD algorithm, both inter- and intra-diversity are greatly improved, especially for large-degree users. Figure \ref{Proportion_ProbeSet} shows the improvements of SMD over MD and Hybrid methods with different size of the probe set. 

\begin{table}[ht]
\caption{Comparisons of four method on two datasets. The length of recommendation list is $L=20$ for computing the metrics related to $L$. Each value is obtained by averaging over ten independent runs.}
\label{compareWithOtherAlgorithms}\centering
\begin{tabular}{ccccc}
\hline
Friendfeed & MD & HC & Hybrid($\lambda^*$=0.67) & SMD($p^*$=0.71) \\ \hline
\textit{RS} & $0.1064$ & $0.1219$ & ${0.1048}$ & ${0.0948}$ \\
\textit{P} & ${0.0200}$ & $0.0120$ & ${0.0209}$ & $0.0190$ \\
\textit{N} & $63.320$ & ${11.196}$ & ${47.602}$ & $70.217$ \\
\textit{H} & $0.9229$ & ${0.9866}$ & ${0.9628}$ & $0.8874$ \\
\textit{I} & $0.1243$ & ${0.0570}$ & ${0.1097}$ & $0.1271$ \\
\textit{Cov} & $0.6060$ & ${0.7676}$ & ${0.7430}$ & $0.5066$ \\
\textit{C} & $0.8650$ & ${0.6652}$ & ${0.7501}$ & $0.9137$ \\ 
\hline
Epinions & MD & HC & Hybrid($\lambda^*$=0.51) & SMD($p^*$=0.77) \\ \hline
\textit{RS} &  $0.1731$ &$0.2179$  & ${0.1642}$ & ${0.1696}$ \\
\textit{P} & ${0.0208}$ &$0.0075$  & ${0.0252}$ & $0.0196$ \\
\textit{N} & $242.23$ & ${8.7263}$ & ${151.33}$ & $255.87$ \\
\textit{H} & $0.6403$ & ${0.9831}$ & ${0.8802}$ & $0.5726$ \\
\textit{I} & $0.1302$ & ${0.0352}$ & ${0.1140}$ & $0.1330$ \\
\textit{Cov} & $0.2628$ &${0.6447}$& ${0.6678}$ & $0.1974$ \\
\textit{C} & $0.9778$ & ${0.7671}$ & ${0.8555}$ & $0.9856$ \\ \hline
\end{tabular}
\end{table}

\begin{figure}[htb]
\centering
\begin{tabular}{ccc}
  \includegraphics[scale = 0.33]{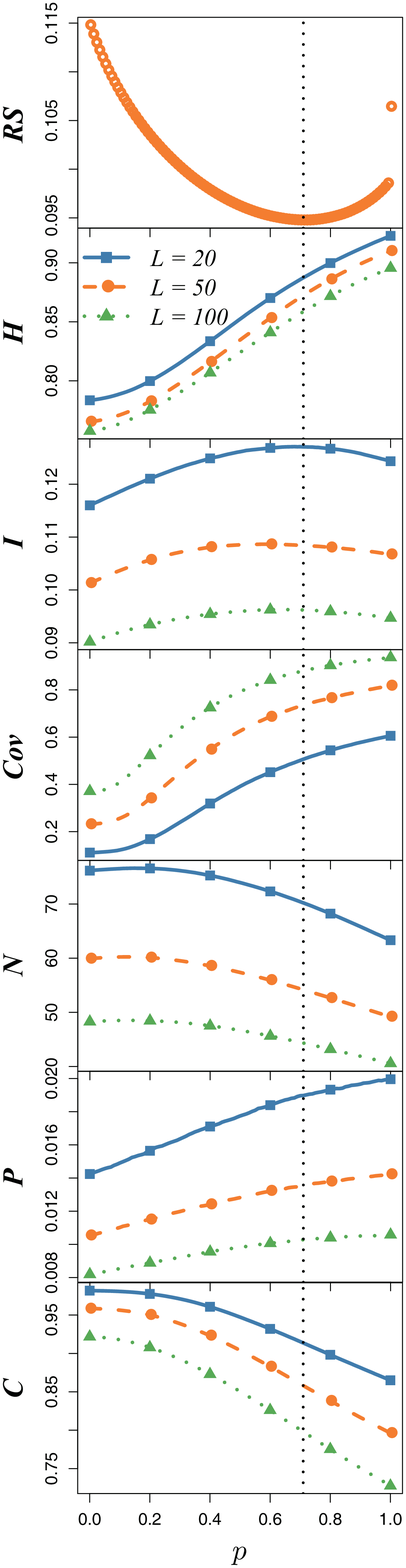}
  &\includegraphics[scale = 0.33]{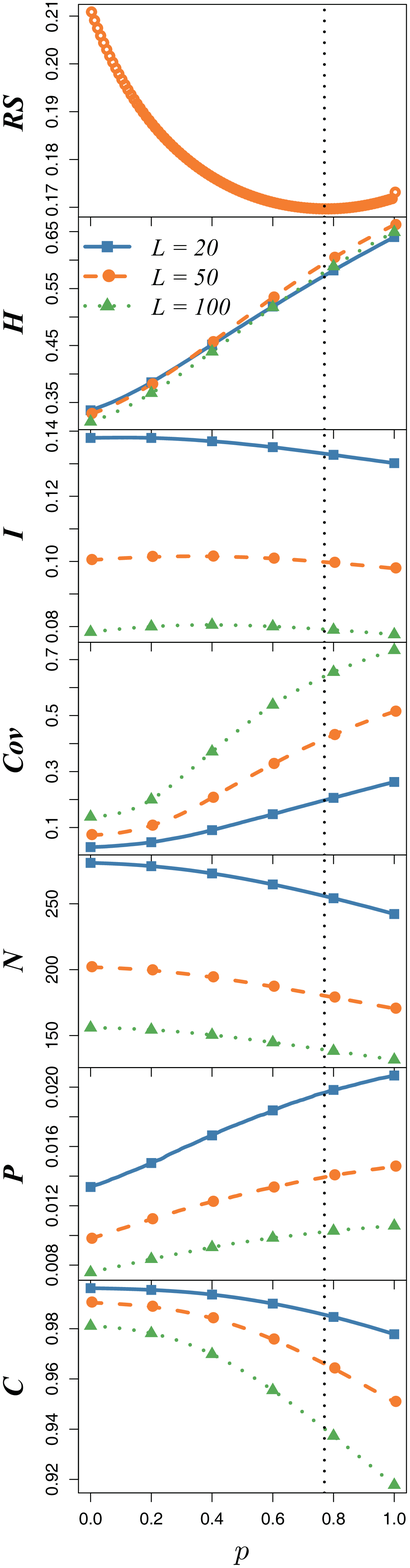}\\
  (a) Friendfeed& (b) Epinions
\end{tabular}
\caption{The performance of SMD method measured by seven metrics under different parameter $p$. The dotted lines are used to mark the optimal parameter $p$, which is $0.71$ in \textit{Friendfeed} and $0.77$ in \textit{Epinions}, respectively.}
\label{ep_ff_all}
\end{figure}

\begin{figure}[tbh]
\centering
\begin{tabular}{ccc}
\includegraphics[width=0.225\textwidth]{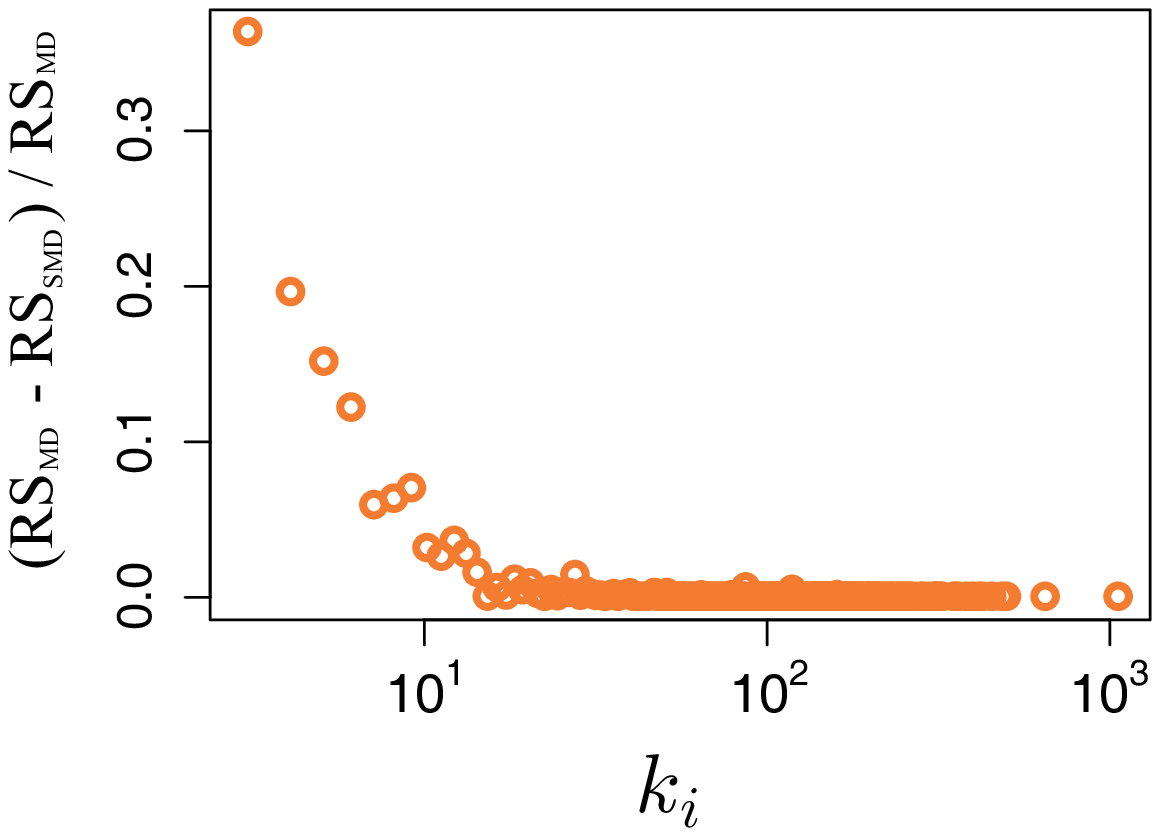} &
\includegraphics[width=0.225\textwidth]{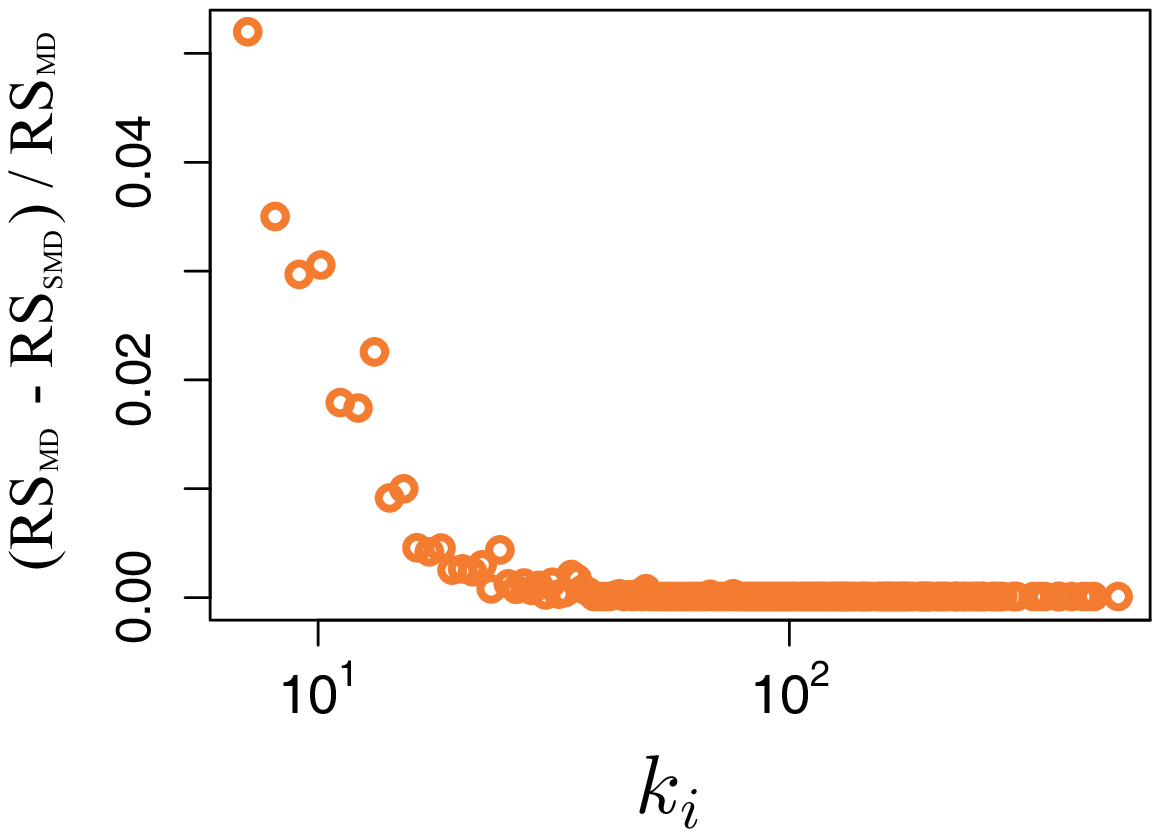} \\
(a) Friendfeed&(b) Epinions\\
\end{tabular}
\caption{The accuracy improvement of SMD with $p=1-\epsilon$ comparing with MD (i.e., $p=1$) for users with different degrees $k_i$.}
\label{ff_jump_99999999_on_1}
\end{figure}

Figure \ref{ep_ff_all} shows the dependence of all metrics on parameter $p$. For \textit{Friendfeed} dataset, \textit{RS} reaches the minimum value of $0.095$ at $p$=$0.71$, corresponding to an improvement of $10.96\%$ over the original MD (the result of SMD at $p=1$). The improvement for the \emph{Epinions} dataset is not significant as that for \emph{Friendfeed}. As shown in figure \ref{ep_ff_all}(b), we can see that the \textit{RS} obtained by SMD improves by only $1.97\%$ compared to that obtained by MD. Besides, we notice that \textit{RS} decreases sharply when $p$ decreases from $p=1$ for both datasets. In order to explain the sharp decrease, we compare the \textit{RS} values of each target user in the cases of $p=1$ and $p=1-\epsilon$, where $\epsilon=1\times 10^{-8}$ is a very small number to show a slight impact of social information on recommendation. From figure \ref{ff_jump_99999999_on_1}, we find that a lower \textit{RS} value is obtained for all target users in the cases of $p=1-\epsilon$ compared to the case of $p=1$, which implies that even a small number of resources spreading on the social network will be beneficial to the recommendation accuracy. In addition, figure \ref{ff_jump_99999999_on_1} also indicates that the smaller the user's degree $k_{i}$, the greater the improvement, implying that the benefit from SMD is more obvious on small-degree users. Comparing with \emph{Epinions}, \emph{Friendfeed} has more inactive users (users with small $k_i$) who need more information from the social network to improve the recommendation accuracy. Therefore the optimal parameter $p^*$ of \emph{Friendfeed} is much smaller than $p^*$ of \emph{Epinions}. We further investigate some inactive users and find that most of their uncollected items assign zero resource by the MD method. Although $\epsilon$ is very small, it breaks the degenerate state by assigning a small number of resources to the relevant items through social network. 

\subsection{Solving the cold-start problem}

\begin{figure}[htb]
\centering
\begin{tabular}{ccc}
\includegraphics[width=0.225\textwidth]{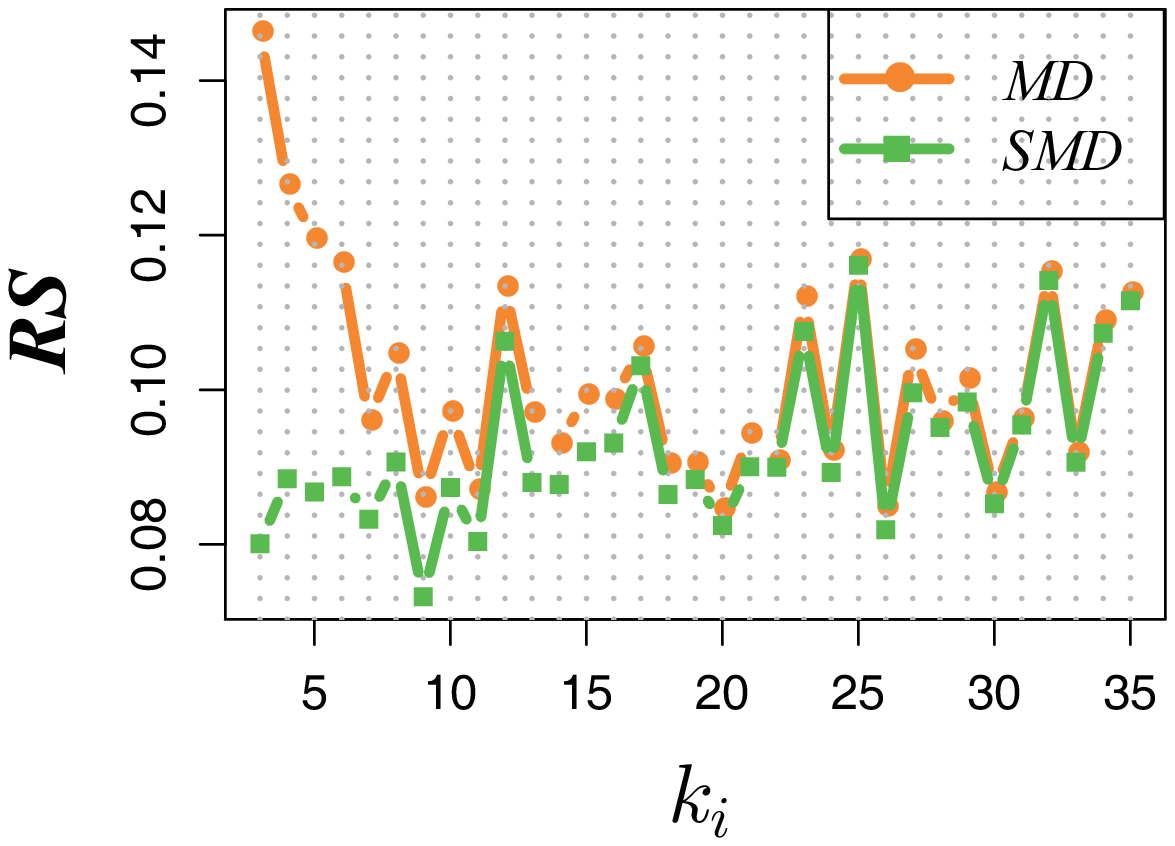} &
\includegraphics[width=0.225\textwidth]{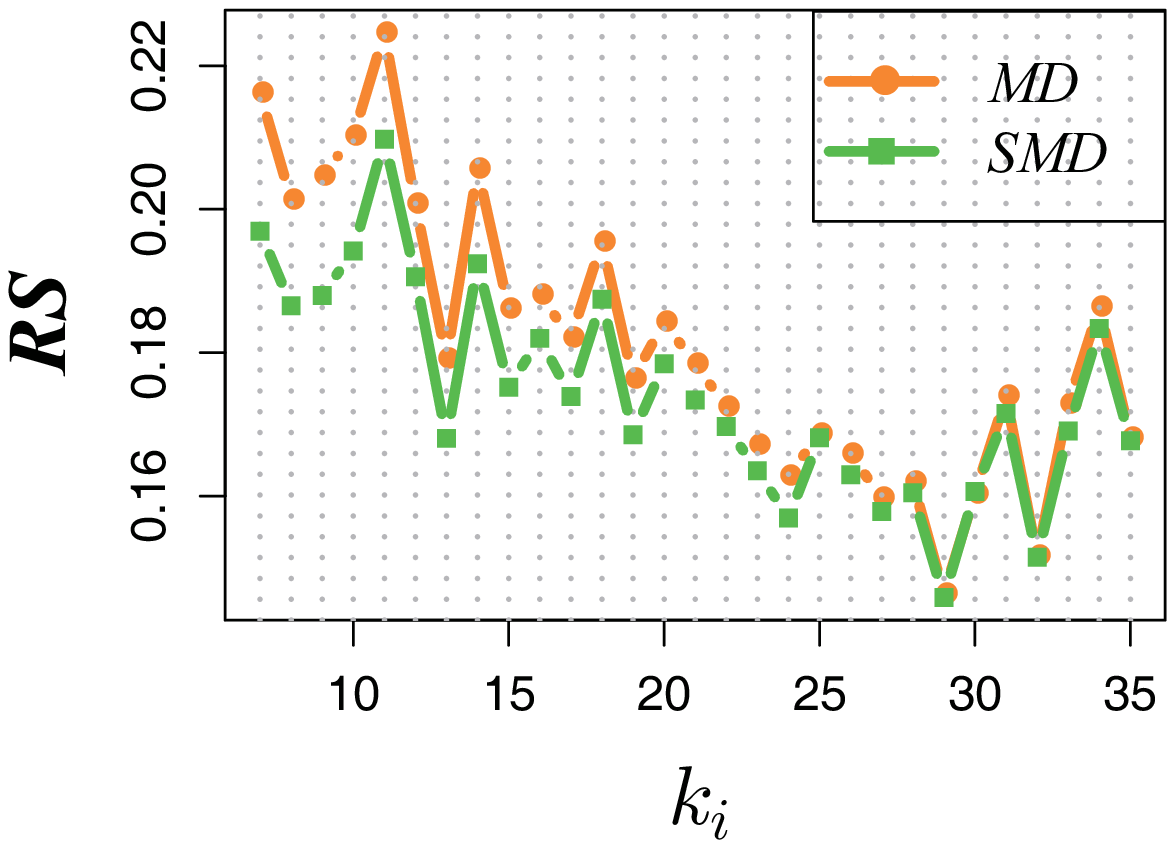} \\
(a) Friendfeed&(b) Epinions\\
\end{tabular}
\caption{ Comparison of the optimal \textit{RS} values of SMD and MD methods. (a) for \textit{Friendfeed}, the \textit{RS} reaches its optimal value at $p^*$=$0.71$; (b) for \textit{Epinions}, the \textit{RS} reaches its optimal value at $p^*$=$0.77$. SMD method outperforms MD method in both datasets.}
\label{degreeMDAndOPT}
\end{figure}

\begin{figure}[htb]
\centering
\begin{tabular}{ccc}
\includegraphics[width=0.225\textwidth]{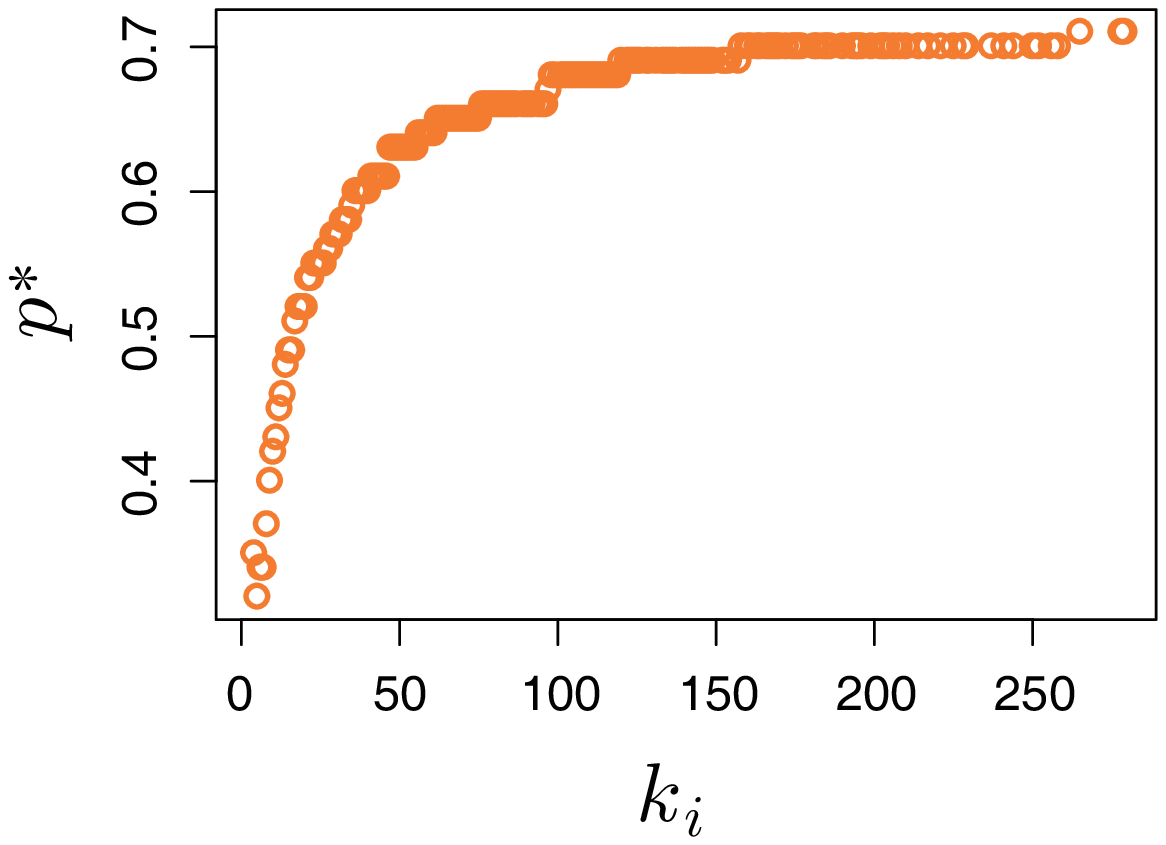} &
\includegraphics[width=0.225\textwidth]{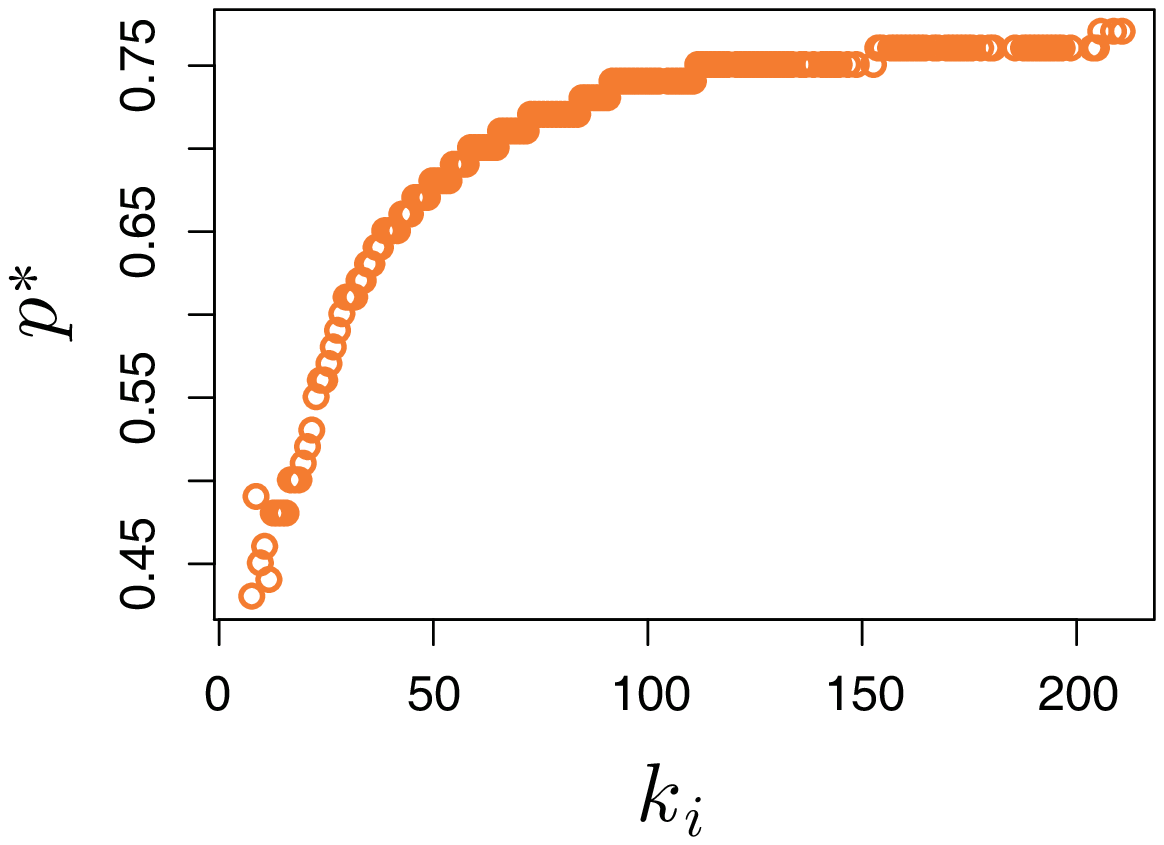} \\
(a) Friendfeed&(b) Epinions\\
\end{tabular}
\caption{The Optimal parameter $p^*$ as a function of user degree $k_i$. For each $k_i$, the averaged $RS$ reaches its optimal value at a $p^*$ value. (a)  \textit{Friendfeed}, $k_i\leq 280$; (b) \textit{Epinions}, $k_i\leq 210$.}
\label{OptimalProbabilityUnderDegree}
\end{figure}

One of the main challenges in recommender systems is the cold-start problem. That is how to give accurate recommendations to new users who have little or even no historical information. Due to insufficient information, the recommendations for these users usually have low accuracy. Experimental results on these two datasets show that our SMD method can largely improve the recommendation accuracy of users with small degrees. We thus focus on users with degree $k_i\le 35$ in \textit{Friendfeed} and \textit{Epinions}. As we can see from figure \ref{degreeMDAndOPT}, the average \textit{RS} of these small-degree users obtained by SMD is all lower than that obtained by MD. We further find a positive correlation between the user's degree $k_i$ and the optimal parameter $p^*$, see figure \ref{OptimalProbabilityUnderDegree}. With the increasing of user's degree $k_i$, the optimal $p^*$ increases faster when $k_i$ is small, and slows down when $k_i$ exceeds the average value. These results suggest that the social information is more useful to improve the recommendation accuracy for small-degree users than for large-degree users.

\begin{figure}[htb]
\centering
\begin{tabular}{ccc}
\includegraphics[width=0.225\textwidth]{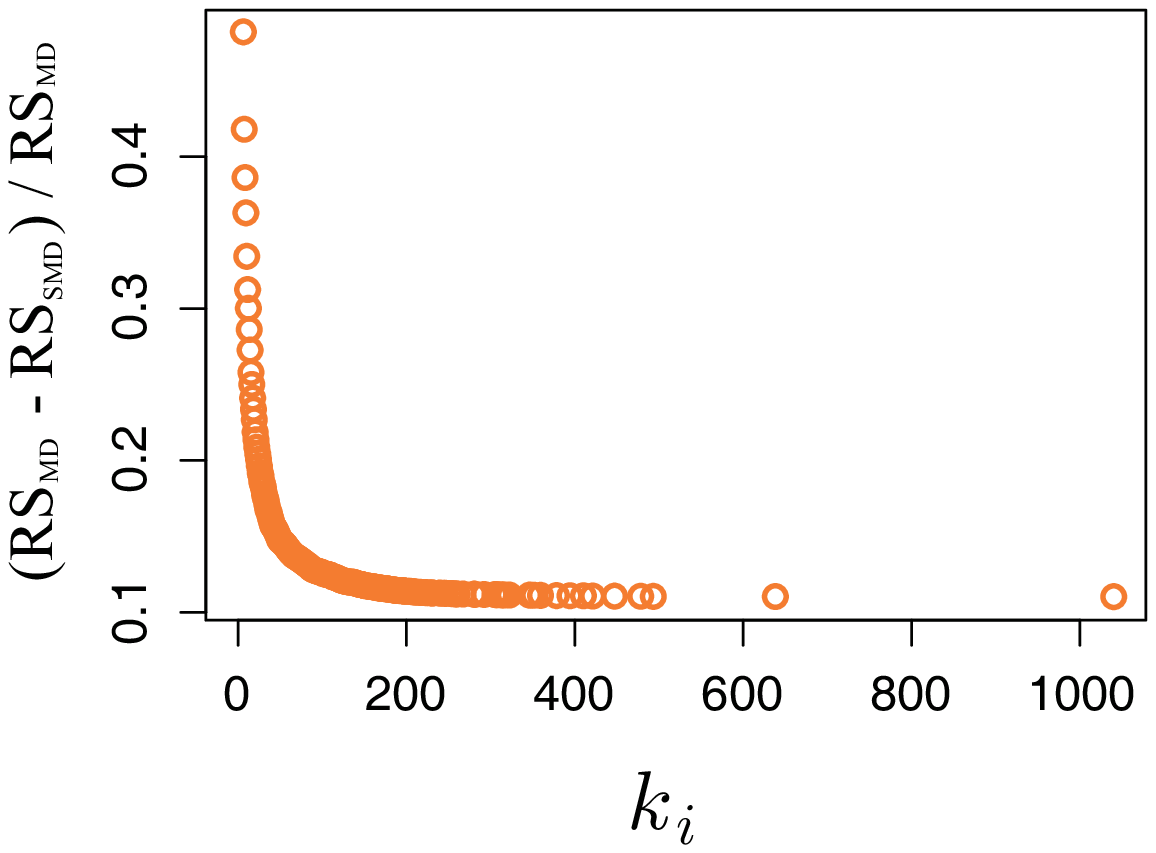} &
\includegraphics[width=0.225\textwidth]{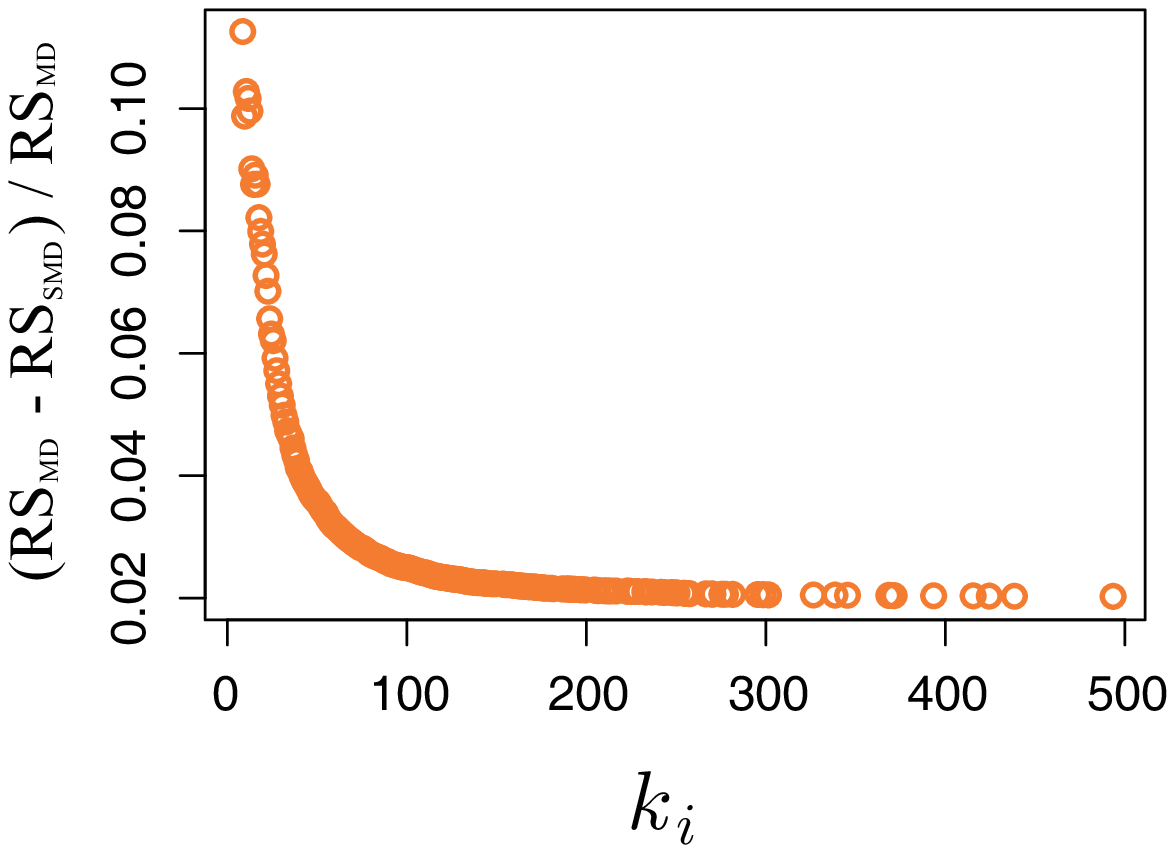} \\
(a) Friendfeed&(b) Epinions\\
\end{tabular}
\caption{Improvement on \textit{RS} by comparing SMD with MD. For each $k_i$, the averaged $RS$ reaches its optimal value at a $p^*$ value, we calculate the improvement of optimal $RS$ to the $RS$ of MD (i.e., $p=1$). }
\label{ImproveMD}
\end{figure}

We show the improvement on \textit{RS} obtained by SMD compared to MD in figure \ref{ImproveMD}, which again suggests that SMD is more beneficial for small-degree users. We further examine those users who collect less than $5$ items in \textit{Friendfeed} and less than $13$ in \textit{Epinions}. They account for more than $20\%$ of the users in these two datasets. We find that the minimum \textit{RS} is achieved at $p=0.34$ in \textit{Friendfeed} and $p$=$0.48$ in \textit{Epinions} and the improvements on \textit{RS} compared with the original MD are $38.56\%$ and $8.74\%$, respectively. Furthermore, the precision can be also improved by $10.81\%$ in \textit{Friendfeed} when the length of recommendation list is 20.Comparisons of other metrics can be found in table \ref{compareWithOtherAlgorithms_InactiveUsers}.

% Inactive users:
\begin{table}[ht]
\caption{The recommendation results for small-degree users. The length of recommendation list is $L=20$ for computing the metrics related to $L$. Each value is obtained by averaging over ten independent runs.}
\label{compareWithOtherAlgorithms_InactiveUsers}\centering
\begin{tabular}{ccccc}
\hline
Friendfeed & MD & HC & Hybrid($\lambda^*$=1) & SMD($p^*$=0.34) \\ \hline
\textit{RS} & $0.1312$ & $0.1519$ & $0.1312$ & ${0.0806}$  \\
\textit{P} & $0.0097$ & $0.0057$ & $0.0097$ & ${0.0108}$ \\
\textit{N} & $46.932$ & ${11.7434}$ & $46.932$ & $69.270$ \\
\textit{H} & ${0.9665}$ & $0.9907$ & ${0.9665}$ & $0.8717$ \\
\textit{I} & $0.0949$ & ${0.0537}$ & $0.0949$ & $0.1136$\\
\textit{Cov} & $0.4619$ & ${0.5485}$ & $0.4619$ & $0.2285$\\
\textit{C} & $0.7856$ & ${0.6808}$ & $0.7856$ & $0.9397$\\ 
\hline
Epinions & MD & HC & Hybrid($\lambda^*$=0.67) & SMD($p^*$=0.48) \\ \hline
\textit{RS} & $0.2045$ & $0.2515$ & $0.2016$ & ${0.1866}$ \\
\textit{P} & $0.0080$& $0.0025$ & ${0.0089}$ & $0.0075$ \\
\textit{N} & $217.54$& ${8.2413}$ & $140.28$ & $266.41$ \\
\textit{H} & $0.7194$& ${0.9855}$ & $0.8901$ & $0.5054$ \\
\textit{I} & $0.1137$& ${0.0260}$ & $0.0910$ & $0.1342$ \\
\textit{Cov} & $0.2064$ & ${0.4035}$ & $0.4087$ & $0.0875$ \\
\textit{C} & $0.9553$& ${0.7977}$ & $0.8420$ & $0.9885$ \\ \hline

\end{tabular}
\end{table}

\begin{figure}[tbh]
%\centering
\includegraphics[scale = 0.6]{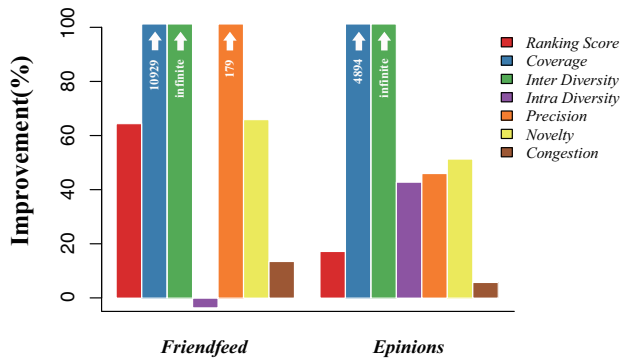}
\caption{The improvement of  recommendation results obtained by SMD comparing with GRM for new coming users. The length of recommendation list is $L=20$. For the sake of illustration, we only present the fractional improvement which is less than $100\%$ by the bar chart, and the values of the fractional improvements which are larger than $100\%$ are labeled on the corresponding bars.}
\label{ff_GlobalRankMD}
\end{figure}

The extreme case for cold-start problem is the recommendation for new coming users who haven't collected any items and thus the recommender systems don't have any information about their favors and is not able to suggest relevant items for them. In this case, a widely used method is called the Global Ranking Method (GRM), with which the items will be recommended to new users according to their popularity. The MD, HC and Hybrid methods are all fail for new coming users, while our SMD method can give better recommendations than GRM. To demonstrate the effectiveness of SMD in solving the cold-start problem, we consider the group of users with $k_i<4$ for \textit{FriendFeed} and $k_i<8$ for \textit{Epinions}, and move all his/her existing user-item links as the probe set while keep the rest of the network as the training set. We then use SMD method to recommend items for them. As shown in figure \ref{ff_GlobalRankMD}, the recommendation results obtained by applying SMD show an extensive improvement over GRM for a wide range of metrics except intra-user recommendation diversity for \textit{FriendFeed}, and SMD outperforms GRM over all the tested metrics for \textit{Epinions}. Especially, the inter-user recommendation diversity is improved significantly since recommendations are now personalized instead of identical for every user. The above results show that SMD is a good substitute of GRM for providing users with personalized recommendations in the cold-start period. This is particularly important for recommender systems to demonstrate their capability of providing personalized recommendation to new coming users rather than a role of an alternative form of advertisement.

%%%%%%------------------figure--ff_GlobalRankMD--------------------------%%%%%%

\section{Conclusion}

In this paper, the information of social networks has been applied to design personalized recommendation algorithms. A social mass diffusion method (SMD) is proposed by considering the resources allocation process on the integrated network which consists of users' social network and user-item bipartite network. Two benchmark datasets, namely \emph{Friendfeed} and \emph{Epinions}, are used to evaluate the performance of the new method. The results show that SMD can improve the recommendation accuracy comparing with the original mass diffusion method which is based solely on the user-item bipartite network. Especially, for the small-degree users, the improvements are significant. In addition, the information of social relationships can help to infer users' preferences especially when the users have only a few records or even the new users without any historical records. Experimental results show that SMD significantly outperforms the conventional popularity-based algorithms in cold-start period in terms of both accuracy and diversity.

\end{document}